\documentstyle[aps,preprint,eqsecnum]{revtex}
\begin{document}
\draft
\preprint{$\hbox to 5 truecm{\hfill Alberta-Thy-18-94}\atop
{\hbox to 5 truecm{\hfill gr-qc/9405033}\atop
\hbox to 5 truecm{\hfill May 1994}}$}
\title{Quasilocal energy for a Kerr black hole}
\author{Erik A. Martinez
\footnote{electronic address: martinez@phys.ualberta.ca}}
\address{Theoretical Physics Institute,\\
Department of Physics, University of Alberta,\\
Edmonton, Alberta T6G 2J1, CANADA.}
\maketitle
\begin{abstract}
\noindent{\baselineskip=16pt
The quasilocal energy associated with a constant stationary time slice of the
 Kerr spacetime is presented.  The calculations are based on a recent proposal
\cite{by} in which quasilocal energy is derived from the Hamiltonian of
spatially bounded gravitational systems. Three different classes of boundary
surfaces for the Kerr slice are considered (constant radius surfaces,
round spheres, and the ergosurface). Their embeddings in both the Kerr
slice and flat three-dimensional space (required as a normalization of the
energy) are analyzed.  The energy contained within each surface is explicitly
calculated in the slow rotation regime and its properties discussed in detail.
The energy is a positive, monotonically decreasing function of the boundary
surface radius. It approaches the Arnowitt-Deser-Misner (ADM) mass at spatial
infinity and reduces to (twice) the irreducible mass at the horizon of the Kerr
black hole. The expressions possess the correct static limit and include
negative contributions due to gravitational binding. The energy at the
ergosurface is compared with the energies at other surfaces.  Finally, the
difficulties involved in an estimation of the energy in the fast rotation
regime are discussed. \par}
\end{abstract}
\vspace{7mm}
\pacs{PACS numbers: 04.20.Cv, 05.30.Ch, 97.60.Lf}
\vfill
\eject

\section{Introduction}

Even at the classical level, general relativity differs from other physical
theories in that it accepts several alternative `definitions' of quasilocal
energy. Despite considerable efforts, no definite expression for quasilocal
energy has yet appeared.  In fact, different proposals provide
satisfactory definitions of  energy when applied to the appropriate
physical situations. In the present paper we study the quasilocal energy
recently proposed in Ref.\cite{by} as applied to a Kerr black hole \cite{kerr}.

One of the appealing features of the definition of quasilocal energy adopted in
this paper is its straightforward derivation from the gravitational
action for a spatially bounded region \cite{by}.
Consider a spacetime ${\cal M}$ foliated by spacelike
hypersurfaces denoted by $\Sigma$. The spacetime is spatially bounded by the
three-dimensional surface ${^3\!{B}}$, and  the intersection of $\Sigma$ with
the
boundary  ${^3\!{B}}$ is a  two-dimensional surface ${^2\!{B}}$ with
induced two-metric  ${\sigma}_{ab}$. The quasilocal energy of $\Sigma$
contained
within the two-dimensional  boundary ${^2\!{B}}$ arises from the action as the
value
of the Hamiltonian that generates unit time translations in ${^3\!{B}}$
orthogonal to
the surface ${^2\!{B}}$.
It can be expressed as the proper surface integral \cite{by}
\begin{equation}
E = \varepsilon - \varepsilon ^0 \equiv {1\over\kappa} \int_{{^2\!{B}}} d^2x
\sqrt{\sigma}\,
\bigl(k-k^0 \bigr) \ , \label{energy}
\end{equation}
involving the trace $k$ of extrinsic curvature of ${^2\!{B}}$ as embedded in
$\Sigma$. The extrinsic curvature is defined so that
$k$ equals (minus) the expansion of the unit outward-pointing spacelike
normal to ${^2\!{B}}$ in $\Sigma$, and  units are chosen so that $G = c = \hbar
=1$, and $\kappa \equiv 8 \pi$. This expression for the energy also appears in
the study of self-gravitating systems in thermal equilibrium, where it
plays the role of the  thermodynamical  energy  conjugate to
inverse temperature \cite{york86,bmy}. A definition of quasilocal energy
intimately related to the energy (\ref{energy}) has been proposed in Ref.
\cite{israel}. For a discussion  of quasilocal energy proposals, see Refs.
\cite{by,h} and references cited therein.

The expression (\ref{energy})  includes a subtraction term $\varepsilon^0$.
This term reflects the subtraction term proposed in Ref.
\cite{gh} for the gravitational action  and  represents a normalization of the
energy with respect to a reference space.
The reference space is a fixed hypersurface of some fixed spacetime and $k^0$
 the trace of extrinsic curvature of a two-dimensional surface  in this space
whose induced metric is ${\sigma}_{ab}$. The proposal (\ref{energy}) is only
sensible if the metric ${\sigma}_{ab}$ can be embedded uniquely in the
reference space. The reference space is usually assumed to be a flat
three-dimensional
slice $E^3$ of flat spacetime. In this case  several theorems exist \cite{by}
that guarantee that the function $k^0$ is uniquely determined for all
topologically spherical surfaces with positive curvature two-metrics.
To specify the quasilocal energy one requires therefore the intrinsic metric of
the surface ${^2\!{B}}$ and its embedding in  $\Sigma$ as well as the embedding
in the three-dimensional reference space $E^3$ of a surface whose intrinsic
geometry equals that of ${^2\!{B}}$.

The energy (\ref{energy}) has been previously calculated  for  static
spacetimes \cite{by}, where the negative contributions to the  energy
arising from  gravitational binding have been studied.  It is of interest to
extend this analysis to stationary spacetimes, and particularly to Kerr black
holes. Besides its possible astrophysical applications, this analysis is
 necessary in the study of rotating black holes in thermal
equilibrium with a heat bath.  For instance, one of the boundary data
specified
at the boundary surface for the density of states in a microcanonical ensemble
description of  rotating black holes \cite{by2,bmy} is  precisely the
quasilocal
energy (\ref{energy}).

We evaluate in what follows the quasilocal energy (\ref{energy}) for constant
stationary time  Kerr slices $\Sigma$ spatially bounded by three
different types of  boundary surfaces ${^2\!{B}}$ with two-sphere topology.
However, it is not easy to evaluate (\ref{energy}) exactly. Besides the
technical problems arising from the complicated structure of the spacetime, it
is not possible to embed an  arbitrary  two-dimensional boundary surface of the
Kerr space $\Sigma$  in $E^3$  and therefore to calculate
the subtraction term $\varepsilon^0$.  While exact results will be provided
whenever possible, most of the energy calculations  have to be performed in the
slow rotation regime. This regime consists of assuming   $|a| /r \ll 1$ (with
$a$ denoting the specific angular momentum of the black hole and $r$  an
appropriately defined radial distance), but imposes no constraints on the
behavior of the ADM mass $M$ \cite{adm}.
Whilst this approximation cannot give a fair description of the fast rotation
regime, it is nevertheless  physically interesting and allows one to study  the
effects of angular momentum on the quasilocal energy. We discuss in detail the
embedding for all choices of boundary surface in this approximation and
estimate separately the terms $\varepsilon$ and $\varepsilon^0$.

In section II we present general expressions for the proper integrals involved
in the quasilocal energy for arbitrary stationary axisymmetric spacetimes (that
is, without assuming the Kerr metric form). These expressions will be  useful
to calculate not only $\varepsilon$ for different choices of surfaces embedded
in
the Kerr slice $\Sigma$ but also $\varepsilon^0$ for appropriate surfaces
embedded in $E^3$. In section III we consider the quasilocal energy contained
within  two-surfaces ${^2\!{B}}$ defined by  constant value of the
Boyer-Lindquist
radial coordinate. These  surfaces are  natural counterparts (in the case of
non-zero angular momentum)  to  surfaces of constant Schwarzschild radius. They
are naturally adapted to the  coordinates and considerably simplify  the
calculation of the energy. In particular, the outer horizon of the Kerr
spacetime is a surface of this type.  We then present various properties and
limiting values  of the energy for these surfaces and discuss  the
contributions due to gravitational binding.

We turn in section IV to the energy contained within a  round spherical
boundary  of the Kerr slice $\Sigma$. There are two advantages in the use of
round spheres in the present calculation: firstly, a sphere can be
always embedded in flat three-dimensional space and consequently the
subtraction term $\varepsilon^0$ can be easily determined. Secondly, a
spherical surface is particularly useful in the study of black hole
thermodynamics. Besides the quasilocal energy mentioned above, the boundary
data
for rotating black holes in a microcanonical ensemble include
a chemical potential
(associated with the conserved angular momentum) and the two-geometry $\sigma$
of the two-dimensional boundary ${^2\!{B}}$ \cite{bmy,by2}.
The ADM mass $M$  and specific angular momentum $a$  of rotating  black hole
configurations in thermal equilibrium inside the cavity are not free parameters
but have to be determined as functions of those boundary data by inverting the
boundary data equations \cite{york86,bmy}. Though this procedure will not be
discussed here,  it  is greatly simplified whenever the boundary surface is a
sphere; in this case the information about the two-geometry of the boundary
surface is fully contained in the area  of the sphere, which does not depend
explicitly on the parameters $M$ and $a$ \cite{bmy}. The quasilocal energy
expressions obtained in this section are then contrasted in the appropriate
limits with the corresponding expressions obtained in section III.

The quasilocal energy within the ergosurface is calculated in section V. The
result is contrasted with the energy at the horizon evaluated in section III,
with the energy within a $r=2M$ surface, and with the energy within a constant
radius surface whose  area equals the surface area of the ergosurface.
We conclude  with a summary of the results and some comments regarding their
generalization beyond the slow rotation regime.

\section{General expressions}

Consider a four-dimensional stationary axisymmetric spacetime. A
three-dimensional axisymmetric spacelike hypersurface $\Sigma$ of
this  spacetime is described by the line element
 \begin{equation}
 h_{ij} dx^i dx^j = b^2 dy^2 + c^2 d \vartheta ^2 + d^2 \, d \varphi ^2 \ ,
\label{3metric}
 \end{equation}
where $x^i (i = y, \vartheta, \varphi)$ denote arbitrary coordinates adapted to
the symmetry and the
metric functions $b$, $c$, and $d$ depend only on the `radial'  coordinate $y$
and the `azimuthal' coordinate $\vartheta$.
An arbitrary  two-dimensional axisymmetric surface ${^2\!{B}}$ (with the
topology of a two-sphere)  embedded  in the three-dimensional space $\Sigma$ is
defined by the relation  $y = F(\vartheta)$, where $F$ is a function of the
azimuthal angle and the parameters of the solution. Its two-dimensional line
element is of the form
\begin{equation}
\sigma _{ab} dx^a dx^b
= (b^2 {F'}^2 + c^2)  d \vartheta ^2 + d^2 \, d \varphi ^2 \ , \label{2metric}
\end{equation}
where a prime  denotes differentiation with respect to the coordinate
$\vartheta$. The functions $b$, $c$, and $d$ in (\ref{2metric}) are evaluated
at $y = F(\vartheta)$.

Let  $n^{i}$ denote the unit outward-pointing spacelike normal to
${^2\!{B}}$ as embedded in the hypersurface $\Sigma$.
In terms of  metric functions its components are
\begin{equation}
 (n^y, n^{\vartheta}, n^{\varphi}) = {1 \over {\sqrt{c^2 + b^2 F'^2}}} \big(
c/b\, ,  -b\, F'/c\, ,0 \big) \ . \label{normal}
\end{equation}
The extrinsic curvature of the two-surface ${^2\!{B}}$ as embedded in $\Sigma$
is
denoted by $ k_{{\mu\nu}}$. Its trace $k$ represents (minus) the expansion of
the
normal vector $n^i$ and can be written in terms of a Lie derivative with
respect to $n^i$ as $k = -{\cal L}_{n}  \ln \sqrt{h}$. Using (\ref{normal}) and
the notation $\alpha \equiv  c^2 d $,
$\beta \equiv  b^2 d \, F' $,
$\lambda \equiv  c^2 + b^2 F'^2 $, and
$\delta \equiv  \ln \lambda $,
the trace is
\begin{equation}
{k} = -{{1} \over {bcd}} \bigg[ \big( \alpha \,{\lambda}^{-1/2} \big)_{,y} -
\big( \beta \,{\lambda}^{-1/2} \big)_{,\vartheta} \bigg]
\Bigg|_{{\scriptscriptstyle} y=F(\vartheta)} \ , \label{explicitk}
\end{equation}
while its proper surface integral  becomes
\begin{equation}
{1 \over \kappa} \int _{{^2\!{B}}} { k} \,\sqrt{\sigma}\, d\vartheta d\varphi =
-{1 \over 4} \int_{0} ^{\pi} d \vartheta \,{1 \over {bc}} ( \alpha _{,y} -
\beta_{,\vartheta} - {{\alpha}\over 2} \, \delta_{,y} + {{\beta}\over 2}
\delta_{,\vartheta})\bigg| _{{\scriptscriptstyle}  y=F(\vartheta)} \
. \label{int}
\end{equation}
This integral is  evaluated at the surface $y = F(\vartheta)$.
Both integrals in expression (\ref{energy}) are of the
general form (\ref{int}), but each one  involves different values for the
functions $b$, $c$, and $d$ as well as different coordinate definitions.

Consider in particular the  Kerr line element expressed in
Boyer-Lindquist \cite{bl,mtw} coordinates ($t,r,\theta,\phi$) in which the
rotational Killing vector field takes the form  $\xi_\phi ^\mu = (\partial
/{\partial
\phi})^\mu$.  The spacelike hypersurfaces $\Sigma$ are  slices of
constant stationary time $t$. Their line element in  Boyer-Lindquist
coordinates is of the form  (\ref{3metric}) with  $b^2 \equiv \rho ^2 /\Delta$,
$c^2 \equiv \rho ^2$,  and  $d^2 \equiv \Sigma^2 \sin ^2 \theta /{\rho
^2}$.
The Kerr functions $\rho$, $\Delta$, and $\Sigma$ are functions of
position and the parameters $(M,a)$ of the solution, and are explicitly given
 by
\begin{eqnarray}
 \rho ^2 &=& r^2 + a^2 \cos^2 \theta, \nonumber \\
\Delta &=& r^2  - 2Mr + a^2, \nonumber \\
\Sigma ^2 &=& (r^2 + a^2)^2 - \Delta a^2 \sin^2 \theta \ .
\label{kerrf}
\end{eqnarray}
(The metric function $\Sigma ^2$ is not to be confused with the
spacelike hypersurfaces  denoted by $\Sigma$.)  The symbol $M$ denotes the ADM
mass of the Kerr black hole and $a \equiv J/M$ denotes its specific angular
momentum. For non-extremal  black holes $|a| \leq M$. A two-dimensional
surface ${^2\!{B}}$ in $\Sigma$ is described by the equation $r = f(\theta
;M,a)$,
and
its
line element is of the form (\ref{2metric}). It is not difficult to see from
the structure (\ref{kerrf}) of the Kerr metric functions that the evaluation of
$\varepsilon$ for the Kerr slice $\Sigma$ bounded by the surface
$r=f(\theta;M,a)$ is technically involved.

For later purposes we note that the line element of ${^2\!{B}}$ in the slow
rotation
approximation becomes
 \begin{eqnarray}
ds^2 &\approx &\Bigl( 1 + {{a^2}\over{r^2}} \cos^2 \theta \Bigr)
 \Biggl[{\biggl( 1 - {{2M}\over{r}} + {{a^2}\over{r^2}} \biggr)}^{-1}
 {\biggl({{df}\over{d\theta}} \biggr)}^2
+ r^2
 \Biggr] \, d\theta ^2 \nonumber \\
&+& r^2  \Biggl[
1 + {{a^2}\over{r^2}}\biggl(1 + {{2M}\over{r}} \sin ^2 \theta \biggr)
\Biggr]
\sin ^2 \theta \, d\phi ^2 , \label{ksurfslow}
\end{eqnarray}
where $r=f(\theta;M,a)$, and terms of order $O(1/r^4)$ and higher have been
neglected.

\section{A constant radius surface}

We  calculate first the  energy
within the simplest choice of surface ${^2\!{B}}$, defined by
$r= r_0 ={\rm constant}$.  We assume that ${r_0} \geq r_+$, where $r_+$
represents the outer horizon of the Kerr black hole.  Using (\ref{int}), the
integral $\varepsilon$ for the surface $r = {r_0}$ can be written explicitly as
\begin{eqnarray}
\varepsilon  &=& -{{{r_0}}
\over 2} \sqrt{ 1 -{{2M}\over{{r_0}}} + {{a^2}\over{{r_0} ^2}} }\, \times
\nonumber \\
 & &\int_{0} ^{\pi} d\theta \sin \theta { {1 +
{{a^2}\over{{r_0} ^2}} - {{a^2}\over{2 {r_0} ^2}}(1 - {{M}\over{{r_0}}}) \sin^2
\theta
} \over
{\sqrt{ 1 + {{a^2}\over{{r_0} ^2}} \cos ^2 \theta}
\sqrt{ 1 + {{a^2}\over{{r_0} ^2}} (1 + \cos ^2 \theta) +
{{2Ma^2}\over{{r_0} ^3}}\sin^2 \theta  +
{{a^4}\over{{r_0} ^4}} \cos^2 \theta }
}
} \ .
\end{eqnarray}
Unfortunately, this  integral cannot be expressed in terms of simple
functions. Whilst it is possible that the inclusion of the subtraction term
simplifies the total integral (\ref{energy}) for the quasilocal energy, we
turn our attention to the slow rotation regime defined by $|a|/{r_0} \ll 1$.
In this approximation the  integral becomes
\begin{eqnarray}
\varepsilon  &=&
-{{{r_0}} \over 2} \sqrt{ 1 - {{2M}\over{{r_0}}} + {{a^2}\over{{r_0} ^2}} }
\times\nonumber \\
& &\int_{0} ^{\pi} d\theta \sin \theta \Bigg[
1 - {{a^2}\over{2{r_0} ^2}} \bigg({{M}\over{{r_0}}} + \cos ^2 \theta
\bigg( 1 - {{M}\over{{r_0}}}\bigg) \bigg) + O \bigg({{a^4}\over{{r_0}
^4}}\bigg)
\Bigg] \ ,
\end{eqnarray}
where only leading order terms in the expansion around the small parameter
$|a|/{r_0}$ are considered. This is easily integrated in terms of polynomials
with the result
\begin{equation}
\varepsilon = -{r_0} \sqrt{ 1 - {{2M}\over{{r_0}}} + {{a^2}\over{{r_0} ^2}} }
 \, \,
\Bigg[
1 - {{a^2} \over { 6 {r_0} ^2}} \bigg( 1 + {{2M}\over{{r_0}}} \bigg) +
 O \bigg({{a^4}\over{{r_0} ^4}} \bigg)
\Bigg] \ . \label{e1}
\end{equation}
It is important to note that no approximations have been made inside
the square-root appearing in $\varepsilon$ (in fact,  the square-root times
${r_0}$ is equal to $\Delta ^{1/2}$). For large
values of $r_0$,  $\varepsilon \to M-r_0$. The presence of the divergent term
indicates the need of a subtraction term for the energy, which we consider
below.

In order to calculate the subtraction term $\varepsilon^0$ it is necessary
to find a two-dimensional surface immersed in flat three-dimensional
space $E^3$ whose intrinsic geometry equals the intrinsic geometry of
the $r={r_0}$ surface in the Kerr space. Once this surface is known, the trace
$k^0$ of its extrinsic curvature as embedded in $E^3$ will allow us to
calculate $\varepsilon^0$.
Since the intrinsic geometry of a two-dimensional  surface is
completely characterized by its scalar curvature, we can obtain the equation
for the surface by requiring its scalar curvature to be  equal
to the scalar curvature of the $r={r_0}$ surface in the Kerr space.
Due to the complicated structure of the Kerr
spacetime, the construction will  be confined to the  regime $|a|/{r_0} \ll 1$.

Using (\ref{ksurfslow}), the intrinsic line element
of the $r={r_0}$ surface in the Kerr space  is
\begin{equation}
ds^2 = {r_0}^2 \,\biggl( 1 + {{a^2}\over{{r_0}^2}} \cos^2 \theta \biggr)
  \, d\theta ^2
+{r_0}^2  \sin^2 \theta \biggl( 1 + {{a^2}\over{{r_0}^2}} + {{2Ma^2}\over
{{r_0}^3}}
\sin ^2 \theta + O(a^4/{r_0}^4) \biggr)
d\phi ^2 , \label{r0metric}
\end{equation}
and its scalar curvature is readily calculated to be
\begin{equation}
\Re  = {2\over{{r_0}^2}} \bigg[
1 + {{a^2}\over{{r_0}^2}} \Big( -2 \cos^2 \theta + {{2M}\over{{r_0}}}
(1-4\cos^2 \theta) \Big) + O(a^4/{r_0}^4)\bigg]\ . \label{scr0}
\end{equation}
In our approximation regime the curvature (\ref{scr0}) is positive for
all values of the azimuthal angle $\theta$ and the surface can be embedded
globally  in three-dimensional Euclidean space $E^3$.  (It is well known
\cite{smarr} that the Kerr horizon cannot be embedded globally in $E^3$
whenever $a^2 > {{r_+}^2}/3$, or equivalently, $a > \sqrt{3} M/2$. In this case
part of the surface, centered around the equator, remains embedded in $E^3$
while another part, centered around the polar caps, becomes embedded in flat
pseudo-Euclidean ($PE^3$) space.)

Consider now the  line element
\begin{eqnarray}
ds^2 _{\rm {\scriptscriptstyle} flat} &=& - (dx^0)^2 + dx^2 + dy^2 + dz^2
\nonumber \\
&=& - (dx^0)^2 + d{\cal R}^2 + {\cal R}^2 d\Theta ^2 + {\cal R}^2 \sin^2 \Theta
\, d\Phi^2  \label{flato}
\end{eqnarray}
of a four-dimensional flat  spacetime. The flat coordinates $({\cal R},
\Theta,
\Phi)$ should not be confused with the Boyer-Lindquist coordinates
$(r,\theta,\phi)$. The relationship between both coordinate systems is
important in the construction of the desired two-dimensional surfaces embedded
in flat space and deserves to be discussed. The line element
(\ref{flato}) can be written in terms of Boyer-Lindquist coordinates by
defining
\begin{eqnarray}
 x^0 &=& \int du + dr \nonumber \\
x &=& (r \cos {\tilde \phi} + a \sin
{\tilde \phi}) \sin \theta \nonumber \\
y &=& (r \sin {\tilde \phi} - a \cos {\tilde
\phi}) \sin \theta \nonumber \\
 z &=& r \cos \theta \ ,\label{xyz}
\end{eqnarray}
where the  Kerr-Schild \cite{bl,mtw} coordinates $(u, {\tilde \phi} )$ are
given
by  $du = dt - (r^2 + a^2)/{\Delta} \, \, dr$, and
$d{\tilde \phi} = d\phi - a/{\Delta}\, \, dr$.
(In fact, in terms of these coordinates the Kerr line element can be written
 as the direct sum of two line elements, one of
them being the flat element (\ref{flato}) ). The relationship between
$({\cal R}, \Theta, \Phi)$ and $(r,\theta,\phi)$ can be obtained easily by
noting that the coordinates $(x,y,z)$ satisfy \cite{bl}
\begin{equation}
{{x^2 + y^2}\over{
r^2 + a^2}} + {{z^2}\over{r^2}} = 1 \ , \label{relation}
\end{equation}
and $x^2 + y^2 + z^2 = {\cal R}^2$. In particular, one obtains
${\cal R}^2 = r^2 + a^2 \sin^2 \theta$ and
\begin{equation}
\cos^2 \theta = {{r^2 +
a^2}\over{r^2 + a^2 \cos^2 \Theta}} \cos^2 \Theta \ ,\, \sin^2 \theta =
{{r^2}\over{r^2 + a^2 \cos^2 \Theta}} \sin^2 \Theta \ ,
\end{equation}
which in the  approximation $|a| \ll r$ reduce to
\begin{eqnarray}
\cos^2 \theta &\approx & (1 +a^2 /r^2 \sin^2 \Theta) \cos^2 \Theta \nonumber \\
\sin^2 \theta &\approx & (1 -a^2 /r^2 \cos^2 \Theta) \sin^2 \Theta \ .
\label{thetas}
\end{eqnarray}

A flat three-dimensional slice of the flat spacetime (\ref{flato}) has line
element
\begin{equation}
ds^2 = d{{\cal R}}^2 + {{\cal R}}^2 d{\Theta}^2 +{{\cal R}}^2
\sin ^2 \Theta \, d\Phi^2 \ .
\label{flat}
\end{equation}
The equation for the desired two-dimensional surface in the flat slice is
denoted  by ${\cal R} = g(\Theta)$, where $g$ is a function of the azimuthal
angle $\Theta$ and the parameters $(M,a,{r_0})$ of the surface in $\Sigma$. Its
intrinsic metric is  obtained from (\ref{flat}).  By virtue of (\ref{thetas}),
it is enough in the slow rotation approximation to assume
\begin{equation}
g(\Theta) =  {r_0} \biggl( 1 +  {{a^2}\over{{r_0}^2}}\, {\omega}(\Theta)
+O(a^4/{r_0}^4) \biggr) ,
 \end{equation}
where ${\omega}(\Theta)$, a function of order one, is to be determined. The
scalar
curvature of this surface is
\begin{equation}
\Re ^0 = {2\over{{r_0}^2}} \bigg[
1 - {{a^2}\over{{r_0}^2}} \Big( {\omega}'' + {\omega}'\cot \Theta + 2{\omega}
\Big) + O(a^4/{r_0}^4)\bigg]\ , \label{scflat}
\end{equation}
where a prime denotes differentiation with respect to $\Theta$. By equating the
scalar curvatures (\ref{scr0}) and (\ref{scflat}) and using (\ref{thetas}) we
obtain a differential equation for ${\omega}(\Theta)$, whose solution is
$ {\omega}(\Theta) = -{M/ {r_0}} \, \cos 2\Theta + {1/2}\, \sin ^2 \Theta $.
This implies that the radius of the surface in $E^3$, as a function of the
azimuthal angle is given by
\begin{equation}
{\cal R} = g(\Theta) = {r_0} \Biggl[
1 + {{a^2}\over{{r_0}^2}} \bigg( {1\over 2} \sin ^2 \Theta - {{M}\over{{r_0}}}
\cos
2\Theta \bigg) + O(a^4/{r_0}^4)\Biggr]\ .  \label{surfflat}
\end{equation}
This equation describes a surface in flat Euclidean space $E^3$ whose intrinsic
curvature equals the intrinsic curvature of a $r=r_0$ surface in the Kerr
slice. This surface, as embedded in $E^3$, is plotted in Figure 1 for
particular values of $M$ and $a$ that satisfy the condition $a^2 \ll {r_0}^2$.
It is a distorted sphere with its major axis along the equator.
For fixed $a^2$, the surface becomes more oblate as $M$ increases.
Its intrinsic metric is given by
\begin{equation}
ds^2 \approx  {r_0}^2 \biggl[
1 + {{a^2}\over{{r_0}^2}} \Bigl( \sin ^2\Theta -
{{2M}\over{{r_0}}} \cos 2\Theta \Bigr) \biggr] ( d\Theta ^2 + \sin^2 \Theta \,
d\Phi ^2 )\ ,
  \end{equation}
whereas its area is
\begin{equation}
A = 4\pi {r_0}^2 \Bigg[ 1 +{{2a^2}\over{3 {r_0}^2}} \bigg(1 + {M\over {r_0}}
\bigg) + O(a^4/{r_0}^4)\Bigg] \ .
\end{equation}

The extrinsic curvature $k^0$ of the surface (\ref{surfflat}) as embedded in
the flat space (\ref{flat}), and its proper integral can  be calculated now
using the expressions (\ref{explicitk}) and (\ref{int}) adapted to the
coordinates $({\cal R}, \Theta, \Phi)$. A long but direct computation gives
the desired integral
\begin{equation}
\varepsilon ^0 = {1 \over \kappa} \int _{{^2\!{B}}} k^0 \sqrt{\sigma}\,
d\Theta d\Phi =
- {r_0} \Bigg[ 1 + {{a^2}\over{3 {r_0} ^2}}\bigg( 1 + {M\over {r_0}} \bigg)
+ O(a^4 /{r_0} ^4) \Bigg]
\ . \label{eo1}
\end{equation}

The energy is obtained by subtracting (\ref{eo1}) from (\ref{e1}), with the
result
\begin{eqnarray}
E &=& {r_0} \Bigg[ 1 -
\sqrt{ 1 - {{2M}\over{{r_0}}} + {{a^2}\over{{r_0} ^2}}}\, \, \Bigg] +
{{a^2}\over{6{r_0} }} \Bigg[ 2 + {{2M}\over{{r_0}}} +  \bigg(1 + {2M \over
{{r_0}}}\bigg) \sqrt{ 1 - {{2M}\over{{r_0}}} + {{a^2}\over{{r_0} ^2}}} \,\Bigg]
\nonumber \\
& & + r_0 \, \, O(a^4/{r_0}^4) \,\,\ , \label{E1}
\end{eqnarray}

\noindent where terms of order $r_0 \, O(a^4/{r_0}^4)$ and higher are
considered negligible.
Expression (\ref{E1}) is the quasilocal
gravitational energy of a (slowly rotating) Kerr slice $\Sigma$ spatially
bounded by a surface of constant Boyer-Lindquist radial coordinate $r_0$.
We stress that no restrictions on the mass $M$ have been imposed in the
calculation of the energy expression (\ref{E1}).

Several features of the above expression are noteworthy. In the
asymptotic limit ${r_0} \to \infty$, the energy approaches the ADM
energy $M$. In the limit of zero angular momentum the
quasilocal energy reduces to
\begin{equation}
E = {r_0} - {r_0} \, \sqrt{ 1 - {{2M}\over {r_0}} } \ ,
\label{Eschwarz}
\end{equation}
which, as expected, is the quasilocal energy (within a surface of radius
$r=r_0$)
for a Schwarzschild black hole of ADM mass $M$.
The small mass limit (or equivalently, the large radius limit) of the energy
(\ref{E1}) is obtained by assuming $M \ll {r_0}$ in (\ref{E1}). In this case
\begin{equation}
 E = M + {{M^2}\over{2{r_0}}} + {{M^3}\over{2{r_0}^2}} +
{{5M^4}\over{8{r_0}^3}}  - {{7 M^2 a^2}\over{6{r_0}^3}} +
{r_0} \, O(a^4/{r_0}^4) \ , \label{newton1}
\end{equation}
where terms of order $O(M^4/{r_0}^4)$ and $O(M^2 a^2/{r_0}^4)$  have been
preserved while terms of order $O(a^4/{r_0}^4)$ and higher have been neglected.
The first term in (\ref{newton1})  represents the total energy at spatial
infinity while the second term represents (minus) the Newtonian gravitational
potential energy associated with building a shell of radius ${r_0}$ and mass
$M$
by bringing the individual constituent particles from spatial infinity. As
discussed in Ref. \cite{by} for the static case, the energy $E$ in this
approximation has  the natural interpretation of the total energy associated
with assembling the gravitational system starting from the boundary of radius
${r_0}$, and it reflects indeed the energy within this boundary.  Notice that
no
terms of order  ${r_0}\, O(a^2/{r_0}^2)$ are present in (\ref{newton1}).

It is of interest to invert the energy equation (\ref{E1}) to obtain the
ADM mass $M$ as a function of the energy $E$, the size ${r_0}$ of the cavity,
and the specific  angular momentum $a$. This not only illustrates the
interpretation of the different terms in the energy expression but also, as
mentioned in the introductory paragraphs,
has relevance in the study of black hole stable configurations in a
gravitational microcanonical ensemble.  In the limit of small energy, the
desired equation is
\begin{equation}
M = E \biggr( 1 - {{E}\over{2{r_0}}} \biggl)
+ \, O(a^2 E^2/{{r_0}}^3) \ .
\end{equation}
In the limit $a \to 0$ this  expression is exact for all values of $E$ and
reduces to the expression for the  Schwarzschild ADM mass in terms of
boundary radius and energy \cite{york86}.

Expression (\ref{E1}) can be used to calculate the energy at the horizon
whenever $|a| \ll r_+$ (or equivalently, whenever  $|a| \ll M$). The
two-dimensional outer horizon \cite{mtw} of a Kerr black hole is a topological
(but not a round) sphere of constant coordinate radius  $r_0 = r_+ = M + (M^2
-a^2)^{1/2}$.
In the present approximation, $r_+ = 2M(1 - a^2/4M^2 + O(a^4/M^4))$.
Since  $\varepsilon (r_+) = 0$, the energy becomes
\begin{eqnarray}
E({r_0} = r_+) &=&  r_+ \bigg[
1 + {{a^2}\over{2 {r_+}^2}} + O(a^4/{r_+}^4) \bigg] \nonumber \\
 &=& 2M \bigg[ 1 -
{{a^2}\over{8M^2}} +  O(a^4/{r_+}^4) \bigg]  \ .\label{Ehorizon}
\end{eqnarray}
Recall now that the irreducible mass $M_i$ of a Kerr black
hole \cite{cr} is proportional to the square root of the surface area of the
horizon and can be written as
 \begin{equation}
 M_i = {1\over 2}\Big( {r_+}^2 + a^2 \Big)^{1/2} = {{r_+}\over 2}  \bigg[
1 + {{a^2}\over{2 {r_+}^2}} + O(a^4/{r_+}^4) \bigg] \ . \label{im}
\end{equation}
Therefore the
quasilocal energy (\ref{Ehorizon}) at the horizon   equals  twice the
irreducible mass,
\begin{equation}
 E({r_0} = r_+) = 2M_i [1 + O(a^4/{M_i}^4 )]
\end{equation}
to leading order in the slow rotation regime. This is a very attractive
property of the energy (\ref{E1}),  valid perhaps beyond the
present approximations, and constitutes one of the main results of this paper.

The
quasilocal energy (\ref{E1}) contained within the constant radius surface  is
positive  for all values of  ${r_0} \geq r_+$. This can be  seen directly from
(\ref{E1}), since the two terms of this expression are positive whenever
${r_0}$
is larger than the radius of the outer Kerr horizon.  The behavior of the
quasilocal energy  as a function of $r_0$ is illustrated in Figure 2 for  $|a|
\ll r_+$. Notice that the energy monotonically increases from $E({r_0} =
\infty)
=
M$ to $E({r_0} = r_+) = 2M_i$.  As shown in (\ref{E1}), the quasilocal energy
has
contributions related to the irreducible mass and to rotational energy. The
interaction of the different energy contributions  is non-linear and shows
itself even in the slow rotation approximation.

For later purposes we observe that the energy within a surface of constant
radius ${r_0} = 2M$ is
\begin{equation}
E(r_0 = 2M) = 2M \Biggl( 1 - {{|a|}\over{2M}} + {{a^2}\over {8 M^2}} +
O(a^3/M^3) \Biggr) \ . \label{E2M}
\end{equation}
Notice that   $E(r_0 =r_+) > E(r_0 = 2M)$.

Finally, the quasilocal energy for an extreme Kerr hole for which $M = |a| \ll
r_0$) is
\begin{equation}
E = M + {{M^2}\over {2{r_0}}} + {{M^3}\over {2{r_0}^2}} +
{r_0} \, O(M^4 /{r_0}^4) \ , \label{extreme1}
\end{equation}
in agreement with the small mass limit (\ref{newton1}).

\section{A spherical two-dimensional boundary}

We turn now our attention to the quasilocal energy of
the Kerr space $\Sigma$ contained within a round spherical surface
${^2\!{B}} = S^2$ of curvature radius $R$ and surface area $A =4 \pi R^2$.
The construction of a round sphere in $\Sigma$ and the energy
calculation will be computed in the slow rotation regime $a^2 \ll {R^2}$
supplemented with the additional condition  $M \ll R$. Under this approximation
terms of order $O(M^2 a^2/R^4)$ and $O(M^4 /R^4)$ will be considered while
terms of order $O(a^4/R^4)$ and higher will be neglected.

The equation $r = h(\theta)$  describing $S^2$ in $\Sigma$ can be obtained  by
requiring its scalar curvature to be
\begin{equation}
\Re  = {2 \over {R^2}}, \label{84}
\end{equation}
where $R$ is a positive constant. The function $h$ depends on
the azimuthal angle and the parameters $(M,a,R)$, and in the slow rotation
approximation takes the form
\begin{equation}
 h(\theta) =  R \biggl( 1 +  {{a^2}\over{R^2}}\, {\eta}(\theta) +
O(a^4/R^4) \biggr) , \label{87}
\end{equation}
where ${\eta}(\theta)$ is a function of order one.
Substitution of (\ref{87}) in (\ref{ksurfslow}) gives us a surface line element
whose intrinsic curvature is
\begin{equation}
\Re  = {2\over{R^2}} \bigg[
1 - {{a^2}\over{R^2}} \Big( {\eta}'' + {\eta}' \cot \theta +
2{\eta}  + 2\cos^2 \theta + {{2M}\over R}(4 \cos^2 \theta -1)  \Big) +
O(a^4/R^4)\bigg]\ , \label{scsph}
\end{equation}
where a prime in this section denotes differentiation with respect to
the Boyer-Lindquist coordinate $\theta$.
By equating the
scalar curvatures (\ref{scsph}) with the scalar
curvature  (\ref{84}) of a round sphere one obtains a differential equation for
${\eta} (\theta)$, whose solution is
$
{\eta}(\theta) = {M\over R} \cos 2\theta - {1\over 2}\sin ^2 \theta
$.
Therefore, a two-dimensional round sphere $S^2$ of curvature radius $R$ in a
slowly rotating Kerr space $\Sigma$ is described by the equation
\begin{equation}
r = h(\theta) = R \biggl[
1 - {{a^2}\over{2R^2}}\sin ^2 \theta
+ {{a^2 M}\over{R^3}}  \cos 2 \theta + O(a^4/R^4) \biggr]. \label{sph}
\end{equation}
As expected from the behavior of the Boyer-Lindquist coordinates, the
coordinate radius of the sphere $S^2$ is
larger at the poles than at the equator.
Its line element in the current approximation is
\begin{eqnarray}
ds^2 &=& R^2 \Biggl[ 1 + {{a^2}\over{R^2}} \biggl( 1 + {{2M}\over
R}\biggr) \cos 2\theta + O(a^4/R^4) \Biggr]d\theta ^2 \nonumber \\
&+& R^2 \sin ^2 \theta \Biggl[ 1 +
{{a^2}\over{R^2}} \biggl( 1 + {{2M}\over R}\biggr) \cos ^2\theta  + O(a^4/R^4)
\Biggr] d\phi ^2 \ . \label{95}
\end{eqnarray}
It can  easily be verified that the area of this surface is $A=4 \pi R^2$.

The quasilocal energy integral $\varepsilon$ can now be calculated directly
from the general  expression (\ref{int}) with $b^2 = \rho ^2/\Delta$,  $ c^2 =
\rho ^2$, and $d^2 = \Sigma ^2 \sin ^2 \theta /\rho ^2$,  and by replacing the
coordinates $(y, \vartheta, \varphi)$ by $(r, \theta, \phi)$. The Kerr metric
functions $\rho^2$, $\Delta$, and $\Sigma ^2$, evaluated at the
spherical surface (\ref{sph}), are
\begin{eqnarray}
\rho ^2 (\theta)&=& R^2 \Biggl[ 1 + {{a^2}\over{R^2}}(2 \cos ^2 \theta - 1)
+ {{2a^2 M}\over{R^3}}(2 \cos ^2 \theta - 1)  + O(a^4/R^4)  \Biggr],\nonumber
\\
\Delta (\theta)&=& R^2 \Biggl[ 1 - {{2M}\over R} +
{{a^2}\over{R^2}} \cos^2 \theta +
{{a^2 M}\over{R^3}}(3 \cos ^2 \theta - 1)
+{{2a^2 M^2}\over{R^4}}(1 -2 \cos ^2 \theta)  + O(a^4/R^4) \Biggr], \nonumber
\\
\Sigma ^2 (\theta) &=& R^4 \Biggr[ 1 + {{a^2}\over{R^2}}
(3\cos^2 \theta -1) +  {{2a^2 M}\over{R^3}}(3 \cos ^2 \theta - 1) + O(a^4/R^4)
\Biggr] \ ,
\label{Ksph}
\end{eqnarray}
in the present approximation.
The integrand becomes a function of $(M, a, R)$ and the coordinate
$\theta$. A long but direct calculation gives the total integral
\begin{equation}
\varepsilon = -R \bigg( 1 - {{2M}\over{R}} \bigg)^{1/2}
\Bigg[ 1 + {{M^2 a^2}\over{R^4}} \bigg( 1 - {{2M}\over{R}} \bigg)^{-1} +
O(a^4/R^4) \Bigg]\ .\label{esph}
\end{equation}
As expected, this expression is divergent for large $R$.  The subtraction term
is easily  calculated in the present case since a two-dimensional sphere  can
always be embedded in  $E^3$. A straightforward calculation gives $k^0 = -2/R$
and
\begin{equation}
{\varepsilon}^0 = -R \ .\label{e00sphere}
\end{equation}
By subtracting (\ref{e00sphere}) from (\ref{esph}) one obtains the
total energy, which can be  written in our approximation as
\begin{equation}
E = M + {{M^2}\over{2R}} + {{M^3}\over{2R^2}} + {{5M^4}\over{8R^3}} -
{{M^2 a^2}\over{R^3}} + \, R \, O(a^4/R^4) \ .\label{Esph}
\end{equation}
This expression gives the quasilocal gravitational energy of a Kerr
space $\Sigma$ (with specific angular momentum  $a^2 \ll R^2$ and mass $M \ll
R$) spatially bounded by a round sphere of surface area $A = 4 \pi R^2$.
Observe that the  specific angular momentum $a$ shows itself only at fourth
order in $E$.

We discuss now some properties of the expression (\ref{Esph}).
The energy tends to $M$ as $R$ tends to infinity and in the zero angular
momentum limit one recovers the energy expression (\ref{Eschwarz}) for  a
Schwarzschild black hole (this limit is expected since  a two-surface of
constant radius $r_0$ in the Schwarzschild spacetime is also a round sphere.)
Observe from expression (\ref{esph}) that
a round spherical surface of area $A = 4 \pi R^2$ surrounding a (slowly)
rotating Kerr black hole of  mass $M \ll R$ contains less energy than a round
spherical surface of the same area surrounding a Schwarzschild black hole of
mass $M$. Figure 3 illustrates the behavior of the energy (\ref{Esph}) as a
function of the quantity $R$ whenever $M \ll R$.

It is interesting to contrast the values of quasilocal energy of $\Sigma$
within
different surfaces. Comparing (\ref{Esph}) with (\ref{newton1}), we observe
that
the energy  contained within a round sphere of curvature radius
$R =l_0$ is  larger than the energy contained within a topologically spherical
surface of Boyer-Lindquist radius $r_0 =l_0$  whenever $l_0$ is large (that is,
$|a| \ll l_0$, $M \ll l_0$).
We can compare also the energy  associated with a round sphere of  area
$A=4\pi R^2$ with the energy associated with a constant radius boundary of the
same surface area. Under our approximation the two areas coincide if
\begin{equation}
R \approx
{r_0} \Bigg( 1 + {{a^2}\over{3{r_0}^2}} \bigg( 1 + {M\over {r_0}} \bigg) \Bigg)
\ . \end{equation}
As can be seen by substituting this in (\ref{Esph}) and comparing the resulting
energy  with  (\ref{newton1}),  the energy within the sphere equals the
energy within the constant radius surface in the present approximation
whenever $R$ and $r_0$ are large. This is not surprising since  the relative
distortion between the constant radius surface and the sphere is very small for
large radii.

\section{Ergosurface}
The timelike limit surface (or ergosurface) of a Kerr black hole is
the boundary of the region in which particles travelling on a timelike curve
can remain on an orbit of the Killing vector field $\xi_t ^\mu =  (\partial
/\partial t)^\mu $ (and
so remain at rest with respect to spatial infinity)\cite{mtw}. The ergosurface
is a topologically spherical surface described by  $r = r_* \equiv M + (M^2 -
a^2 \cos ^2 \theta )^{1/2}$; $r_*$ coincides with the outer horizon radius
$r_+$ at the poles and  equals  $2M$ at the equator. Since the ergosurface is
neither a constant radius surface nor a round sphere, expressions
(\ref{E1}) and (\ref{Esph}) cannot be used to estimate the quasilocal energy
within it. However, the  energy can be evaluated in the slow rotation
approximation $|a| \ll M$. The
term $\varepsilon$ can in fact be evaluated directly from (\ref{int}) for
arbitrary values of angular momentum. The result will not be presented here
because,  as for previous surfaces, the integral cannot be written in terms
of simple functions. However, in the small
rotation regime  the integral can be evaluated explicitly, with the result
\begin{equation}
\varepsilon = -{{ \pi |a|} \over 4}\, \bigg(
1 - {{9 a^2}\over {64 M^2}} + O(a^4/M^4) \bigg) \ . \label{eergo}
\end{equation}

In order to calculate $\varepsilon^0$ we need to construct a surface ${\cal R}
= l(\Theta;M,a)$ embedded in $E^3$ possessing the same intrinsic geometry as
the ergosurface.  Since in the slow rotation approximation $r_* \approx
2M(1-a^2/{4M^2} \cos^2 \theta)$,  the intrinsic metric of the ergosurface is,
to leading order
\begin{equation}
ds^2 = 4M^2
 \Biggl[ 1  + {{3a^2}\over{4M^2}}\cos^2 \theta  \Biggr] \, d\theta ^2 +4M^2
\Biggl[
1  + {{a^2}\over{2M^2}}\bigg(1 -{3\over 2}\cos^2 \theta \bigg)
\Biggr]
\sin ^2 \theta \, d\phi ^2 , \label{ergo}
\end{equation}
while its scalar curvature is
\begin{equation}
\Re = {1\over{2M^2}} \bigg( 1 + {{3a^2}\over{4M^2}} (1 - 6 \cos^2
\theta) + O(a^4/{M^4})\bigg) \ . \label{scergo}
\end{equation}
 By equating (\ref{scergo}) with the  scalar curvature of a surface in flat
space,  we can obtain the desired surface ${\cal R} = l(\Theta;M,a)$ in the
regime $|a| \ll M$. A direct calculation gives
 \begin{equation}
l(\Theta;M,a) = 2M\Bigg[ 1 + {{3a^2}\over{4M^2}}\bigg( 1 - {3\over 2} \cos^2
\Theta \bigg) + O(a^4/M^4)\Bigg] \ .\label{ergoflat}
\end{equation}
Notice that the coordinate radius of this surface in $E^3$ is larger than
$2M$ at the equator and smaller than $2M(1-a^2/4M^2)$ at the poles.   The  line
element of this surface is
 \begin{equation}
ds^2 = 4M^2
 \Biggl[ 1  + {{3a^2}\over{2M^2}}\bigg(1 - {3\over 2} \cos^2 \Theta \bigg)  +
O(a^4/{M^4})  \Biggr] \big( d \Theta ^2 + \sin^2 \Theta \, d\Phi ^2 \big) ,
\end{equation}
while its area is
\begin{equation}
A = 16 \pi M^2 \big(1 + {3a^2}/{4M^2} + O(a^4/{M^4})\big) \ .
\end{equation}
The extrinsic curvature $k^0$ of the surface (\ref{ergoflat}) as embedded  in
flat space provides the subtraction term
\begin{equation}
\varepsilon^0 = -2M \bigg( 1 + {{3a^2}\over{8M^2}} + O(a^4/{M^4}) \bigg)
\ . \label{e0ergo}
\end{equation}

The energy can be obtained now by subtracting (\ref{e0ergo}) from
(\ref{eergo}),
with the result
\begin{equation}
E = \varepsilon - \varepsilon^0 = 2M - {{\pi |a|} \over {4}}
 + {{3a^2}\over{4M}} + {{9\pi |a|^3} \over {256 M^2}}
 + M O(a^4/{M^4})\ .\label{Eergo}
\end{equation}
This is the quasilocal gravitational energy (in the approximation $|a| \ll M$)
of a Kerr slice $\Sigma$ spatially bounded by the ergosurface $r=r_*$. The
above
expression shows that the quasilocal energy contained within the ergosurface is
positive in the present approximation.

It is interesting to calculate the difference between the energy
(\ref{Ehorizon}) evaluated  at the horizon and the energy (\ref{Eergo})
evaluated at the ergosurface.  Due to  gravitational binding energy
contributions, the former is larger than the latter, and
\begin{equation}
E(r=r_+)\, - \, E(r=r_*)   =  {{\pi |a|} \over 4}
 - {{a^2}\over{M}} - {{9\pi |a|^3} \over {256 M^2}}
 + M O(a^4/{M^4})\ .
\end{equation}
 This quantity measures the difference  between
(twice) the irreducible mass and the energy contained within the ergosurface.
It is naturally zero when $a=0$.

The surface of coordinate radius $r=2M$ and the ergosurface $r=r_*$ touch at
the
equator but are well-separated elsewhere, with the former surface enclosing the
latter.  The energy within the ergosurface is larger than the energy
(\ref{E2M})
within the surface $r=2M$, and
\begin{equation}
E(r=r_*)\, - \, E(r=2M) = |a| (1-\pi /4) + a^2/{2M} + M O(a^3/{M^3})
 \ .
\end{equation}

To conclude, observe that  the area of the constant radius
surface ${r_0} = r_c \equiv 2M(1+a^2/4M^2)$ equals the area
of the ergosurface in the approximation $|a| \ll M$. (The former surface is
located outside the horizon since, in this approximation, $r_+ \approx
2M(1-a^2/4M^2)$.)
{}From (\ref{E1}) the quasilocal energy at this surface is
\begin{equation}
E(r_c)
= 2M \Bigg( 1 - {{|a|}\over{\sqrt{2}M}} + {{3a^2}\over{8M^2}} +
O(a^3/M^3)
\Bigg)\ . \label{erc}
\end{equation}
A comparison with (\ref{Eergo}) shows that the energy within the
ergosurface is larger than the energy (\ref{erc}) contained within the
$r=r_c$ surface.

\section{Summary}

We have calculated the quasilocal energy (\ref{energy}) for the constant
stationary time slice of the Kerr spacetime contained within three different
classes of boundary surfaces, namely, constant radius surfaces, round spheres,
and the ergosurface. These surfaces were chosen because they play a special
role in the physics of Kerr black holes.  The calculations were confined to the
slow rotation regime $|a| \ll r$. Unless explicitly stated, the mass $M$ was
not restricted by any approximation. Energy expressions in the small mass
(large
radius) limit were also presented. For surfaces close to the outer horizon of
the Kerr black hole, the quasilocal energy was explicitly calculated under the
assumption that $|a| \ll r_+$ (or equivalently, $|a| \ll M$). Under these
approximations, the curvature of the surfaces involved was positive everywhere
and the embedding in $E^3$ could be done explicitly.  In particular, the
quasilocal energy within  constant radius surfaces and spherical surfaces is a
monotonically decreasing function of the radius. The energy within the
ergosurface is larger than $M$, and the energy at the horizon equals twice the
irreducible mass of the black hole. These results constitute a natural
extension to stationary black holes of previously known results concerning
quasilocal energy for static black holes.

Short of an exact evaluation of (\ref{energy}), it is perhaps not unreasonable
to look for a quasilocal energy expression (say, for $r={r_0}$ surfaces) that
satisfies the four conditions: (1) $E \to M$ as $r\to \infty$, (2) $E \to 2M_i$
as $r\to r_+$, (3) $E \to {r_0} - {r_0}(1-2M/{r_0})^{1/2}$ as $a \to 0$, and
(4) $
E\to {\rm Eqn. (\ref{E1})}$ in the limit $|a| \ll {r_0}$. Whereas it is easy to
find expressions that fulfill conditions (1)-(3), it is difficult to satisfy
(4). In particular, relations of the form $ {r_0} - {r_0}\,
(1-{r_+}/{r_0})^{1/2}$ or
$ {r_0} - {r_0} \, (1-2M_i/{r_0})^{1/2}$ do not satisfy the above criteria.
In any
case, the desired expression for  the quasilocal energy of the Kerr space
$\Sigma$ has to reflect the peculiarities of the embedding of boundary surfaces
in $E^3$ in the fast rotation regime. We hope to return to this issue
elsewhere.

\acknowledgments

It is a pleasure  to thank Werner Israel for his encouragement and for his
critical reading of the manuscript. Research support was received from  the
Natural Science and Engineering Research Council of Canada through a  Canada
International Fellowship.

\vfill\eject
\centerline {\bf FIGURE CAPTIONS}
\smallskip
\parindent=0pt \parshape=3 0truein 6truein .7truein 5.3truein .7truein
5.3truein
Figure 1:~ Embedding diagram (in three-dimensional Euclidean space $E^3$) of  a
two-dimensional surface whose intrinsic geometry coincides with the intrinsic
geometry of a $r=r_0$ surface in the Kerr space $\Sigma$. The figure
corresponds to particular values of $M$ and $a$ that satisfy the condition $a^2
\ll {r_0}^2$. For fixed $a^2$, the surface becomes more oblate as $M$
increases.

\smallskip
\parindent=0pt \parshape=3 0truein 6truein .7truein 5.3truein .7truein
5.3truein
Figure 2:~ The quasilocal energy  as a function of the coordinate radius
${r_0}$ of ${^2\!{B}}$ in the case $|a| \ll r_+$.

\smallskip
\parindent=0pt \parshape=3 0truein 6truein .7truein 5.3truein .7truein
5.3truein
Figure 3:~ The quasilocal energy  as a function of the curvature
radius $R$ of a round spherical surface  for the case $M = 1$, $|a| = .2$.
The energy  approaches $M$ at  spatial infinity.

\vfill\eject

\end{document}